\documentclass[a4paper,reqno,12pt]{amsart}
\usepackage{amssymb,eucal,bbold}
\usepackage{array}
\DeclareMathOperator{\Tr}{\mathrm{Tr}}
\DeclareMathOperator{\im}{\mathrm{Im}}
\DeclareMathOperator{\re}{\mathrm{Re}}

\DeclareMathOperator{\Lie}{\mathrm{Lie}}
\DeclareMathOperator{\dvol}{\mathrm{dvol}}

\newcommand{\Cl}{\mathrm{C}\ell}
\renewcommand{\O}{\mathrm{O}}
\newcommand{\SO}{\mathrm{SO}}
\renewcommand{\Sp}{\mathrm{Sp}}
\newcommand{\Spin}{\mathrm{Spin}}
\newcommand{\1}{\mathbb{1}}
\renewcommand{\AA}{\mathbb{A}}
\newcommand{\CC}{\mathbb{C}}
\newcommand{\HH}{\mathbb{H}}
\newcommand{\OO}{\mathbb{O}}
\newcommand{\RR}{\mathbb{R}}
\newcommand{\so}{\mathfrak{so}}
\newcommand{\eM}{\mathcal{M}}
\newcommand{\eA}{\mathcal{A}}
\newcommand{\eG}{\mathcal{G}}
\begin{document}

\title{Gauge theory and the division algebras}
\author{Jos\'e M Figueroa-O'Farrill}
\address{\begin{flushright}Department of Physics\\
Queen Mary and Westfield College\\
University of London\\
London E1 4NS, UK\end{flushright}}
\email{j.m.figueroa@qmw.ac.uk}
\thanks{Supported by the EPSRC under contract GR/K57824.}
\begin{abstract}
We present a novel formulation of the instanton equations in
8-dimensional Yang--Mills theory.  This formulation reveals these
equations as the last member of a series of gauge-theoretical
equations associated with the real division algebras, including
flatness in dimension 2 and (anti-)self-duality in 4.  Using this
formulation we prove that (in flat space) these equations can be
understood in terms of moment maps on the space of connections and the
moduli space of solutions is obtained via a generalised symplectic
quotient: a K\"ahler quotient in dimension 2, a hyperk\"ahler quotient
in dimension 4 and an octonionic K\"ahler quotient in dimension 8.
One can extend these equations to curved space: whereas the
2-dimensional equations make sense on any surface, and the
4-dimensional equations make sense on an arbitrary oriented manifold,
the 8-dimensional equations only make sense for manifolds whose
holonomy is contained in $\Spin(7)$.  The interpretation of the
equations in terms of moment maps further constraints the manifolds:
the surface must be orientable, the 4-manifold must be hyperk\"ahler
and the 8-manifold must be flat.
\end{abstract}
\maketitle

\section{Introduction}

Gauge theory in higher than four dimensions is rapidly coming of age.
Recent developments in superstring theory, particularly related to the
Matrix Conjecture of \cite{Matrix}, point to the existence of
supersymmetric quantum gauge theories in dimensions where
traditionally we would have expected none to exist: five and six
dimensions so far, but possibly higher \cite{MatrixT4T5,SeibergT5}.
Even more recent work \cite{ChrisYM} also suggests that
higher-dimensional instantons \cite{CDFN,Ward} dominate
certain regimes in the moduli space of M-theory.  In addition, these
higher-dimensional instantons are intimately linked to supersymmetry
\cite{BKS,AOS,BKS2,BTESYM,AFOS,FIM,FKS2} and to the geometry of
riemannian manifolds of special holonomy: Calabi--Yau and
hyperk\"ahler geometries, and especially the exceptional geometries in
seven and eight dimensions \cite{ReyesCarrion,DT,Thomas}.  At the same
time, very little is known about these generalised instantons: very
few solutions are known explicitly \cite{FaNu,FuNi,IP}, and almost
nothing is known about the moduli spaces, although the deformation
complexes are elliptic and formulae for the virtual dimensions can be
obtained \cite{RCPhD,ReyesCarrion}.  This result notwithstanding, the
equality between the virtual dimension and the dimension of the moduli
space (at least at irreducible points) hinges on the vanishing of the
higher cohomology of the deformation complex---a question which has
yet to be addressed.

Judging by the 4-dimensional case, instanton moduli space has a rich
geometry worthy of study on its own right.  It is likely that a
similarly rich geometry will emerge out of the study of the moduli
spaces of higher-dimensional instantons.  This note is a first step in
this direction.  We focus on the generalised instantons in eight
dimensions, proving that they fit inside a family of gauge-theoretical
solitons associated with the division algebras $\CC$, $\HH$ and $\OO$,
and including the flat connections in dimension 2 ($\CC$) and the
(anti-)self-dual connections in dimension 4 ($\HH$).  From this fact,
and by analogy with well-known results in the lower dimensions, we
establish some facts concerning the moduli space of octonionic
instantons.  Among other things, we exhibit the moduli space of
octonionic instantons on a flat 8-dimensional manifold, as an
infinite-dimensional octonionic K\"ahler quotient.

This note is organised as follows.  In Section 2 we discuss the family
of instanton equations in $\RR^N$ associated to the division algebras
$\CC$ (for $N{=}2$), $\HH$ (for $N{=}4$) and $\OO$ (for $N{=}8$).  To
the best of our knowledge, this formulation of the octonionic
instanton equations is novel and has the advantage of exhibiting these
equations as the last member of a well-established sequence.  Using
this reformulation, we show in Section 3 that the instanton equations
can be obtained as the zero loci of generalised moment maps and that
the moduli spaces of instantons can be understood as a generalised
symplectic quotient.  This is of course well known in the complex and
quaternionic case.  In Section 4 we investigate the extension of these
results to more general riemannian manifolds.  This will single out
8-manifolds of $\Spin(7)$ holonomy as those admitting the
8-dimensional instanton equations, and flat 8-dimensional manifolds as
those for which the instanton moduli space can be interpreted as an
octonionic K\"ahler quotient.  Section 5 contains some conclusions and
the paper ends with an appendix on octonionic geometry.

\section{Instanton equations in $\RR^N$}

In this section we introduce the ``instanton'' equations on $\RR^N$
where $N=2,4,8$.  These equations consist in setting to zero the
imaginary part of the Yang--Mills curvature in a way that we will make
precise.  In dimension 2 this equation makes the connection flat, in
dimension 4 (anti-)self-dual, and in dimension 8 it becomes the
octonionic instanton equation introduced in \cite{CDFN}.

\subsection{$\CC$-instantons on $\RR^2$}

A gauge field on $\RR^2$ has components $A_\mu(x)$ for $\mu=1,2$.  It
is convenient to consider complex-valued gauge fields $A(x) = A_1(x) i
+ A_2(x)$.   Multiplication by $i$ defines a $2\times 2$ real matrix
$I$ as follows:
\begin{equation*}
i A(x) = I_{1\mu}A_\mu(x) i + I_{2\mu}A_\mu(x)~.
\end{equation*}
Explicitly we see that $I$ is given by
\begin{equation*}
I = \begin{pmatrix}
    0 & 1\\ -1 & 0
    \end{pmatrix}~.
\end{equation*}
We say that $A_\mu(x)$ is a {\em $\CC$-instanton\/} if its curvature
$F_{\mu\nu}$ satisfies
\begin{equation}\label{eq:Cinstanton}
I \cdot F(x)  \equiv I_{\mu\nu} F_{\mu\nu}(x) = 0~.
\end{equation}
From the explicit form of $I$ we see that $\CC$-instantons are
nothing but flat connections: $F_{\mu\nu}(x) = 0$.

\subsection{$\HH$-instantons on $\RR^4$}

Gauge fields $A_\mu(x)$ in $\RR^4$ can be thought of as
quaternion-valued:
\begin{equation*}
A(x) = A_\mu(x) q_\mu = A_1(x) i + A_2(x) j + A_3(x) k + A_4(x)~,
\end{equation*}
where we have introduced a basis $q_\mu = \{i,j,k,1\}$ for the
quaternion units.  Left multiplication by the imaginary units defines
real $4\times 4$ matrices $I$, $J$ and $K$ as before:
\begin{align*}
i A(x) &= I_{\mu\nu} A_{\nu}(x) q_\mu~,\\
j A(x) &= J_{\mu\nu} A_{\nu}(x) q_\mu~,\\
k A(x) &= K_{\mu\nu} A_{\nu}(x) q_\mu~.
\end{align*}
Explicitly, we have
\begin{equation*}
I = \begin{pmatrix}
    0 & i\sigma_2\\ i\sigma_2 & 0
    \end{pmatrix}~,\quad
J = \begin{pmatrix}
    0 & \1\\ -\1 & 0
    \end{pmatrix}~,\quad\text{and}\quad
K = \begin{pmatrix}
    -i\sigma_2 & 0\\ 0 & i\sigma_2
    \end{pmatrix}~.
\end{equation*}
These matrices obey the quaternion algebra $I^2=J^2=K^2=-\1$ and
$IJ=K$, etc.  They also obey the anti-self-duality equation
\begin{equation*}
I_{\mu\nu} = -\tfrac12 \varepsilon_{\mu\nu\rho\sigma} I_{\rho\sigma}~,
\end{equation*}
and similarly for $J$ and $K$.  We say that $A_\mu(x)$ defines an {\em
$\HH$-instanton\/} if the following equations are satisfied:
\begin{equation}\label{eq:Hinstanton}
I \cdot F(x) = J \cdot F(x) = K\cdot F(x) = 0~.
\end{equation}
This means that $F_{\mu\nu}(x)$ is self-dual:
\begin{equation}\label{eq:selfduality}
F_{\mu\nu} = \tfrac12 \varepsilon_{\mu\nu\rho\sigma} F_{\rho\sigma}~.
\end{equation}
In other words, $A_\mu(x)$ is an instanton in the ordinary sense.

The anti-instanton equations are recovered by considering right
multiplication by the conjugate imaginary units on the quaternionic
gauge field $A(x)$.  This gives rise to matrices $\Tilde I$, $\Tilde
J$ and $\Tilde K$ which are now self-dual.  The matrices are different
because $\HH$ is not commutative.  The analogous equations to
\eqref{eq:Hinstanton} but the tilded matrices, now say that
$F_{\mu\nu}(x)$ is anti-self-dual---in other words, $A_\mu(x)$ is an
anti-instanton.

\subsection{$\OO$-instantons on $\RR^8$}

Let us consider a gauge field $A_\mu(x)$ in $\RR^8$ and turn it into
an octonion-valued field
\begin{equation*}
A(x) = A_\mu(x) o_\mu = A_i(x) o_i + A_8(x)~,
\end{equation*}
where we have introduced a basis $o_\mu$ for $\mu=1,\ldots,8$ for the
octonions such that $o_i$ for $i=1,\ldots,7$ are the imaginary units
and $o_8$ is the identity.  Left multiplication by the imaginary units
$o_i$ gives rise to real $8\times 8$ matrices $I^i$ as follows:
\begin{equation*}
o_i A(x) = I^i_{\mu\nu} A_{\nu}(x) o_\mu~.
\end{equation*}
The matrices $I^i$ cannot satisfy the octonion algebra, because unlike
octonion multiplication, matrix multiplication is associative.
Nevertheless they satisfy the 7-dimensional euclidean Clifford
algebra $\Cl(7)$:
\begin{equation}\label{eq:cl7}
I^i I^j + I^j I^i = - 2\delta_{ij} \1~.
\end{equation}
We define an {\em $\OO$-instanton\/} as a gauge field $A_\mu(x)$
subject to the seven equations
\begin{equation}\label{eq:Oinstanton}
I^i \cdot F(x) = 0~.
\end{equation}
Explicitly, for a particular choice of basis, these seven equations
are given by
\begin{align}\label{eq:8instantonfull}
F_{12} - F_{34} - F_{58} + F_{67} &= 0\notag\\
F_{13} + F_{24} - F_{57} + F_{68} &= 0\notag\\
F_{14} - F_{23} + F_{56} - F_{78} &= 0\notag\\
F_{15} + F_{28} + F_{37} - F_{46} &= 0\\
F_{16} - F_{27} + F_{38} + F_{45} &= 0\notag\\
F_{17} + F_{26} - F_{35} + F_{48} &= 0\notag\\
F_{18} - F_{25} - F_{36} - F_{47} &= 0~.\notag
\end{align}
They can be written in a way analogous to the self-duality equation
\eqref{eq:selfduality}:
\begin{equation}\label{eq:Osd}
F_{\mu\nu} = \tfrac12 \Omega_{\mu\nu\rho\sigma} F_{\rho\sigma}~,
\end{equation}
where $\Omega_{\mu\nu\rho\sigma}$ are the components of a 4-form in
$\RR^8$ given by
\begin{equation}\label{eq:OII}
\Omega = -\tfrac16 I^i \wedge I^i~.
\end{equation}
In fact, equation \eqref{eq:Osd} is the way in which the
octonionic equations are usually presented.

The 4-form $\Omega$ is self-dual, as can be seen by the following
construction.  Octonion multiplication defines a 3-form $\varphi$
in $\RR^7$ by:
\begin{equation*}
o_i\,o_j = -\delta_{ij}\, o_8 + \varphi_{ijk}\, o_k~.
\end{equation*}
Our choice of basis is such that
\begin{equation*}
\varphi = o_{125} + o_{136} + o_{147} - o_{237} + o_{246} - o_{345} +
o_{567}~,
\end{equation*}
where we have used the shorthand $o_{ijk} = o_i \wedge o_j \wedge
o_k$.  We now consider the 7-dimensional Hodge dual of $\varphi$:
\begin{equation*}
\Tilde\varphi \equiv \star_7 \varphi =  o_{1234} - o_{1267} + o_{1357}
- o_{1456} + o_{2356} + o_{2457} + o_{3467}~.
\end{equation*}
Thinking of $\Tilde\varphi$ as a 4-form in $\RR^8$, its
8-dimensional Hodge dual is given by $\varphi\wedge o_8$, whence
we can define a self-dual 4-form $\Omega$ in $\RR^8$ as follows:
\begin{align*}
\Omega & = \Tilde\varphi + \varphi\wedge o_8\notag\\
& = o_{1234} + o_{1258} - o_{1267} + o_{1357} + o_{1368} - o_{1456} +
o_{1478} \notag\\
&\qquad {} + o_{2356} - o_{2378} + o_{2457} + o_{2468} - o_{3458} +
o_{3467} + o_{5678}~.
\end{align*}
This is precisely the 4-form defined in \eqref{eq:OII}.

Interpreting $\RR^8$ as the vector representation of $\SO(8)$, the
4-form $\Omega$ is left invariant by a $\Spin(7)$ subgroup of
$\SO(8)$, one under which the vector representation remains
irreducible.  There are three conjugacy classes of $\Spin(7)$
subgroups in $\Spin(8)$, which are related by triality.  Each of these
subgroups are maximal and they can be distinguished by which one of
the three 8-dimensional irreducible representations of $\Spin(8)$
they split.  Two of these subgroups, call them $\Spin(7)_\pm$, are
subgroups of $\SO(8)$: they leave the vector representation
irreducible, but split one of the two spinor representations.  Let
$\Spin(7)_+$ be the one leaving invariant the 4-form $\Omega$ in
\eqref{eq:OII}.  There is a similar set of equations to the
$\OO$-instanton equations but using instead the 4-form
$\Tilde\Omega$ which is invariant by $\Spin(7)_-$.  These equations
are obtained analogously to \eqref{eq:Oinstanton} but using the
matrices $\Tilde I^i$ obtained by right multiplication by the
conjugate imaginary units.  Indeed the 4-form $\Tilde\Omega$ is
given by an equation similar to \eqref{eq:OII} but using the tilded
matrices.

\subsection{Another reformulation}\label{sec:reform}

The instanton equations can be reformulated in yet another way: as the
reality of a laplacian operator.  Let $\{e_\mu\}$ denote generically a
set of units for the division algebra $\AA$, being one of $\CC$, $\HH$
or $\OO$, and let $\{\Bar e_\mu\}$ denote their $\AA$-conjugate.  Let
$N=\dim\AA$ stand for the real dimension of $\AA$.  Then we will
choose our set of units such that $e_N = \1$ and $\{e_i\}_{i=1}^{N-1}$
are imaginary.  Let $D_\mu$ denote the covariant derivative, and let
$D = D_\mu e_\mu$.  We can think of $D$ as acting on $\AA$-valued
fields $\psi$ (with values in some unitary representation of the gauge
group).  Given two such $\AA$-valued fields $\psi,\phi$ we define
their inner product as
\begin{equation*}
(\psi,\phi) = \int_{\RR^N} \dvol\,\Tr\re \psi^\dagger \phi~,
\end{equation*}
where $N$ is the real dimension of $\AA$, $\Tr$ means the gauge
invariant inner product and ${}^\dagger$ involves conjugation in $\AA$
as well as in the representation of the gauge group.  Let $D^\dagger$
denote the formal adjoint of $D$ relative to this inner product:
\begin{equation*}
D^\dagger = - D_\mu \Bar e_\mu~.
\end{equation*}
It follows immediately from the first of the two identities
\begin{equation}\label{eq:Oids}
e_\mu \Bar e_\nu = \delta_{\mu\nu} \1 + I^k_{\mu\nu}
e_k~\qquad\text{and}\qquad
\Bar e_\mu e_\nu = \delta_{\mu\nu} \1 + \Tilde I^k_{\mu\nu} e_k
\end{equation}
that
\begin{equation*}
D^\dagger D = -D^2 \1 + e_k I^k \cdot F(A)~,
\end{equation*}
where $D^2 = D_\mu D_\mu$.  Therefore we see that the $\AA$-instanton
equations \eqref{eq:Cinstanton}, \eqref{eq:Hinstanton} and
\eqref{eq:Oinstanton} are equivalent to
\begin{equation*}
\im \left(D^\dagger D\right) = 0\quad\text{or equivalently}\quad
D^\dagger D = - D^2 \1~.
\end{equation*}
Similarly from the second identity in \eqref{eq:Oids}, it follows that
\begin{equation*}
D D^\dagger = -D^2 \1 + e_k \Tilde I^k \cdot F(A)~,
\end{equation*}
whence the ``anti-self-duality'' equations are equivalent to the
opposite equation:
\begin{equation*}
\im \left(D D^\dagger\right) = 0\quad\text{or equivalently}\quad
D D^\dagger = - D^2 \1~.
\end{equation*}

\section{Instantons and moment maps}

In this section we show that the instanton equations
\eqref{eq:Cinstanton}, \eqref{eq:Hinstanton} and \eqref{eq:Oinstanton}
can be understood as the zeroes of moment maps associated to the
gauge transformations on the space of connections.  This will prove
that the moduli space of instantons can be seen in each case as a
generalised symplectic quotient: a K\"ahler quotient in the complex
case, a hyperk\"ahler quotient in the quaternionic case, and an
octonionic K\"ahler quotient in the octonionic case.  To the best of
our knowledge, this latter quotient construction is new.

As before we let $\AA$ be any one of the division algebras $\CC$,
$\HH$ or $\OO$, and let $N = \dim\AA$ be its real dimension.  Let us
denote by $\eA_\AA$ the space of $\AA$-valued gauge fields on $\RR^N$.
$\eA_\AA$ is an infinite-dimensional affine space modelled on the
space of Lie-algebra valued 1-forms on $\RR^N$.  $\eA_\AA$ inherits
some geometric structure: it is K\"ahler for $\AA=\CC$, hyperk\"ahler
for $\AA=\HH$ and octonionic K\"ahler (see below) for $\AA=\OO$.  The
group of gauge transformations leaves these structures invariant and
will give rise to moment maps whose components are nothing but the
$\AA$-instanton equations.  As a result the moduli space $\eM_\AA$ of
$\AA$-instantons can be understood as a generalised symplectic
quotient of $\eA_\AA$.  This is of course well-known for $\AA=\CC$ and
$\AA=\HH$.  In what follows we will treat all three cases
simultaneously.

\subsection{$\eA_\AA$ as an infinite-dimensional $\AA$-K\"ahler space}

We will use the following notation: $A(x)$ is a Lie algebra- and
$\AA$-valued gauge field on $\RR^N$.  We will let $\overline{A(x)}$
denote its $\AA$-conjugate.  We will let $\Tr$ denote the invariant
metric on the Lie algebra.  The tangent space to the space $\eA_\AA$
of connections is the space of Lie algebra- and $\AA$-valued
1-forms.  Let $\delta_1 A(x)$ and $\delta_2 A(x)$ be two such
tangent vectors.  As above we will let $e_\mu$ denote a basis for the
$\AA$-units, with $e_N$ being the identity and $e_i$ for
$i=1,\ldots,N{-}1$ being imaginary.  Then if $z = z_\mu e_\mu \in\AA$
with $z_\mu \in \RR$, we can define the following bilinear form:
\begin{equation*}
\langle\!\langle \delta_1 A, \delta_2 A\rangle\!\rangle_z =
\int_{\RR^N} \dvol\, \Tr \re z\, \delta_1 A \,\overline{\delta_2 A}~.
\end{equation*}
Expanding this out, we have
\begin{equation*}
\langle\!\langle \delta_1 A, \delta_2 A\rangle\!\rangle_z =
z_i \omega^i(\delta_1 A, \delta_2 A) + z_N g(\delta_1 A, \delta_2 A)~,
\end{equation*}
where the metric $g$ is defined by
\begin{equation}\label{eq:l2metric}
g(\delta_1 A, \delta_2 A) = \int_{\RR^N} \dvol\, \Tr \re \delta_1 A
\,\overline{\delta_2 A}~,
\end{equation}
and the $N{-}1$ 2-forms $\omega^i$ by
\begin{equation}\label{eq:2forms}
\omega^i(\delta_1 A, \delta_2 A) = \int_{\RR^N} \dvol\, \Tr\re
e_i\,\delta_1 A \,\overline{\delta_2 A}~.
\end{equation}
In components, we have
\begin{equation*}
g(\delta_1 A, \delta_2 A) = \int_{\RR^N} \dvol\, \Tr \delta_1 A_\mu
\,\delta_2 A_\mu~,
\end{equation*}
and
\begin{equation*}
\omega^i(\delta_1 A, \delta_2 A) = \int_{\RR^N} \dvol\, I^i_{\mu\nu}
\Tr \delta_1 A_\nu \, \delta_2 A_\mu~.
\end{equation*}
It is then easy to see that the metric is indeed symmetric and that
the 2-forms are antisymmetric.  Moreover both $g$ and $\omega^i$ are
constant (i.e., do not depend on the connection $A(x)$ on which they
are defined) and hence covariantly constant relative to the
Levi-Civita connection.  We see that $\eA_\AA$ is therefore K\"ahler
for $\AA=\CC$, hyperk\"ahler for $\AA=\HH$ and octonionic K\"ahler for
$\AA=\OO$.  If we do not need to specify $\AA$, we will simply say
that $\eA_\AA$ is {\em $\AA$-K\"ahler\/}.  Notice that $\HH$-K\"ahler
is {\em not\/} quaternionic K\"ahler but hyperk\"ahler.  We use the
term octonionic K\"ahler in a rather narrow sense explained in the
appendix.

\subsection{The $\AA$-valued moment map}

The group of gauge transformations acts by conjugation on the tangent
vectors $\delta A$ and commute with the action of $\AA$.  Because
$\Tr$ is pointwise invariant under conjugation, we see that both the
metric and the K\"ahler forms are gauge invariant.  Let us analyse
more closely the invariance of the K\"ahler forms under infinitesimal
gauge transformations; that is, under $\delta A = D\epsilon$, where $D
= e_\mu D_\mu$ is the $\AA$-valued covariant derivative and $\epsilon$
is a Lie algebra valued function on $\RR^N$.  Taking our cue from the
finite-dimensional case, when the Lie derivative along a vector field
$v$ of a closed 2-form $\omega$ is zero, the contraction
$\imath(v)\omega$ is locally exact, whence there exists (at least
locally) a function $\Phi(v)$ so that $\imath(v)\omega = d\Phi(v)$.  The
functions $\Phi(v)$ are the components of the moment map.

In our case, we have that the contraction of the closed 2-form
$\omega^i$ with the infinitesimal gauge transformation $D\epsilon$ is
given by
\begin{align*}
\omega^i(D\epsilon,\delta A) &= \int_{\RR^N} \dvol\, \Tr\re
e_i\, D\epsilon \,\overline{\delta A}\\
&= \int_{\RR^N} \dvol\, I^i_{\mu\nu} \Tr D_\nu\epsilon\,\delta A_\mu\\
&= \int_{\RR^N} \dvol\, I^i_{\mu\nu} \Tr \epsilon\,D_\mu\delta A_\nu\\
&= \int_{\RR^N} \dvol\, I^i_{\mu\nu} \Tr \epsilon\,\delta F_{\mu\nu}~.
\end{align*}
In other words, the components of the moment map are
\begin{equation*}
\Phi^i(\epsilon) = \int_{\RR^N} \dvol\, \Tr \epsilon I^i\cdot F~.
\end{equation*}
The moment map itself is given by
\begin{equation*}
\Phi^i = I^i \cdot F(A)~,
\end{equation*}
which can be thought of as a map from the space $\eA_\AA$ of
connections to the dual of the Lie algebra $\Lie(\eG)$ of the group of
gauge transformations.  Acting on an element $\epsilon$ in
$\Lie(\eG)$, we obtain $\Phi^i(\epsilon)$.  The zero locus of the
moment map $\Phi^i$ are then the connections for which $I^i\cdot F =
0$.

Furthermore the moment map, as a function: $\Phi^i: \eA_\AA \to
\Lie(\eG)^*$, is equivariant under the infinitesimal 
action of $\eG$, acting on $\eA_\AA$ as infinitesimal gauge
transformations and on $\Lie(\eG)^*$ as the coadjoint representation.
To see this notice that, if $\epsilon,\eta\in\Lie(\eG)$, then
\begin{align*}
\delta_\epsilon \Phi^i (\eta) &= 
\int_{\RR^N} \dvol \Tr \eta\, I^i \cdot \delta_\epsilon F(A) \\
&= \int_{\RR^N} \dvol \Tr \eta\, I^i\cdot [F,\epsilon] \\
&= \int_{\RR^N} \dvol \Tr [\epsilon,\eta]\, I^i \cdot F \\
&= \Phi^i([\epsilon,\eta])~.
\end{align*}
This means that the zero locus of moment map $\Phi^i$ is preserved by
$\eG$ and we can consider the orbit space.  Let $\eA^0_\AA \subset
\eA_\AA$ denote the set of connections $A$ for which $\Phi^i = 0$
for all $i$.  This is nothing but the space of $\AA$-instantons,
whence the orbit space $\eA^0_\AA/\eG$ is then the moduli space
$\eM_\AA$.  In other words, we have proven that the moduli space
$\eM_\AA$ of $\AA$-instantons on $\RR^N$ is an infinite-dimensional
$\AA$-K\"ahler quotient of the space $\eA_\AA$ of connections.  It is
well-known that in the case $\AA = \CC$ (respectively, $\AA=\HH$) the moduli
space inherits the structure of a K\"ahler (respectively,
hyperk\"ahler) manifold.  It is not hard to show that this persists in
the octonionic case.  Details will appear elsewhere.



\section{Instantons on riemannian manifolds}

In this section we investigate whether the instanton equations
\eqref{eq:Cinstanton}, \eqref{eq:Hinstanton} and \eqref{eq:Oinstanton}
make sense on manifolds other than $\RR^2$, $\RR^4$ and $\RR^8$
respectively, and whether the interpretation in terms of moment maps
persists.  We will see that although the complex and quaternionic 
instantons make sense on any (orientable) riemannian manifold of the
right dimension, the octonionic equations only make sense in a
manifold whose holonomy is contained in $\Spin(7)$.  Moreover the
interpretation of the $\HH$-instanton equations in terms of moment
maps will force the manifold to be hyperk\"ahler, whereas for the
$\OO$-instanton it will force it to be flat.

\subsection{The instanton equations on riemannian manifolds}
\label{sec:mfds}

In order for the $\AA$-instanton equations to make sense on an
arbitrary manifold, it is necessary that the structure group of the
tangent bundle preserve the subbundle of 2-forms which define the
equations.  We will take all our manifolds to be riemannian, so that
the group of the tangent bundle reduces to $\O(N)$.  Any further
reduction of the structure group can then be understood as a reduction
of the holonomy of the metric.

In the 2-dimensional case, the bundle of 2-forms is a line bundle,
hence under a change of coordinates the 2-form $I$ will always go
back to a multiple of itself.  Therefore the $\CC$-instanton equation
makes sense on any 2-dimensional manifold.

In four dimensions, the 2-forms $I$, $J$, and $K$ are a local basis
for the anti-self-dual 2-forms.  The maximal subgroup of $\O(4)$
which respects the split $\bigwedge^2 = \bigwedge^2_+ \oplus
\bigwedge^2_-$ into self-dual and anti-self-dual 2-forms is
$\SO(4)$, whence provided that the manifold is orientable, the
$\HH$-instanton equations make sense.  This can also be understood
from the alternate form \eqref{eq:selfduality} of the $\HH$-instanton
equations: we now need that the volume form
$\varepsilon_{\mu\nu\rho\sigma}$ be covariantly constant, which means
that the holonomy group must be $\SO(4)$.

In eight dimensions we obtain a stronger restriction on the manifold.
The holonomy group must respect the split $\bigwedge^2 = \bigwedge^2_7
\oplus \bigwedge^2_{21}$, where $\bigwedge^2_7$ is the subbundle
spanned by the $I^i$ and $\bigwedge^2_{21}$ is its orthogonal
complement.  This latter subbundle is spanned by the antisymmetric
products $I^iI^j - I^jI^i$ and hence corresponds to the Lie algebra
$\so(7)$.   The above split is the eigenspace decomposition of the map
$\bigwedge^2 \to \bigwedge^2$ defined by $\omega \mapsto \star (\Omega
\wedge \omega)$ with $\Omega$ defined by \eqref{eq:OII}.  The maximal
subgroup of $\SO(8)$ which preserves $\Omega$, and hence the above
split, is $\Spin(7)_+$. Therefore the holonomy of the metric has to be
contained in this $\Spin(7)$ subgroup.

\subsection{Moment maps for instantons on riemannian manifolds}

Finally we investigate the persistence of the interpretation of the
$\AA$-instanton equations as the zero locus of a moment map in the
space $\eA_\AA$ of connections, and hence of the moduli space as an
infinite-dimensional $\AA$-K\"ahler quotient.

For this to be the case, we have to endow $\eA_\AA$ with the structure
of an infinite-dimensional $\AA$-K\"ahler manifold.  It is not hard to
show that now it is no longer sufficient to preserve the subbundle of
2-forms spanned by the $I^i$ but that each $I^i$ must be invariant
under the holonomy group.  In two dimensions this constrains the
surface to be K\"ahler, which is simply the condition that it be
orientable.  In four dimensions, the fact that $I$, $J$, and $K$ are
constant under the holonomy group, trivialises the bundle of
anti-self-dual forms.  The holonomy must then be contained in one of
the $\Sp(1)$ factors in $\SO(4)$; in other words, the manifold must be
hyperk\"ahler.  Finally, in eight dimensions the fact that the $I^i$
are parallel, means that the manifold is OK, which as discussed in
the Appendix, implies that it is flat.  We summarise this results in
the following table.

\begin{table}[h!]
\centering
\setlength{\extrarowheight}{2pt}
\begin{tabular}{|c|c|c|c|}
\hline
& & \multicolumn{2}{c|}{($\dim\AA$)-manifolds admitting}\\
$\AA$ & $\AA$-instanton & instanton equation & quotient construction\\
\hline
$\CC$ & $F=0$ & arbitrary & orientable\\
$\HH$ & $F = \pm \star F$ & orientable & hyperk\"ahler \\
$\OO$ & $F\in\bigwedge^2_{21}$ & $\Spin(7)$ holonomy & OK ($\implies$ 
flat)\\ \hline
\end{tabular}
\vspace{8pt}
\caption{$\AA$-instanton equations and their allowed manifolds.}
\end{table}

\section{Conclusion}

In this paper we have reformulated the eight-dimensional instanton
equation introduced in \cite{CDFN} in a way that exhibits it naturally
as a member of a family of equations associated to the real division
algebras $\CC$, $\HH$ and $\OO$, and comprising flatness in dimension
2 and self-duality in dimension 4.  The usual way in which the
octonionic equations are presented, namely equation \eqref{eq:Osd},
has the advantage of suggesting generalisations to geometries in which
one has a harmonic 4-form, but at the same time obscures the relative
simplicity of the equations.  Moreover, it does not distinguish the
8-dimensional case from the other ones, and it also treats both
equation \eqref{eq:Osd} and the ``dual'' equation,
\begin{equation}\label{eq:Oasd}
F_{\mu\nu} = -\tfrac16 \Omega_{\mu\nu\rho\sigma} F_{\rho\sigma}~,
\end{equation}
on an equal basis.  In fact, one often finds in the literature that
equations \eqref{eq:Osd} and \eqref{eq:Oasd} are referred to as the
self-duality and anti-self-duality equations respectively.  This
nomenclature suggests a symmetry between these equations which is not
present in the octonionic case since, for example, the spaces have
different dimension.  In our opinion, self-duality and
anti-self-duality correspond to which way the division algebra $\AA$
acts: if on the left or on the right, and are hence related by a
change of orientation on the manifold.  Although there has been some
work in the literature concerning equation \eqref{eq:Oasd}, we believe
this equation not to be as fundamental as \eqref{eq:Osd}.  This can
already be seen not just in the results of the present paper but also,
for example in \cite{AFOS}, where it is shown that supersymmetry
singles out equation \eqref{eq:Osd}.

The original motivation for this paper was to examine the moduli space
of octonionic instantons for $\Spin(7)$ holonomy 8-manifolds.
Alas, we have found that unless the manifold is flat, the octonionic
K\"ahler quotient construction does not work.  Nevertheless the
geometry of the manifold on which the instanton equations are defined
does influence the geometry of the moduli space of instantons.  For
example, the moduli space of instantons on a K\"ahler 4-manifold is
itself K\"ahler, even though it loses its interpretation as a K\"ahler
quotient.  Similarly it is possible to show that if the holonomy of
the 8-manifold is further reduced, say to a subgroup of
$SU(4)\subset\Spin(7)$, then the moduli space inherits a K\"ahler
structure.  In this case, the instanton equations are the celebrated
Donaldson--Uhlenbeck--Yau equations.

In \cite{FKS2} we used supersymmetry to exhibit a relation between the
octonionic instanton moduli space on an 8-manifold $M\times K$,
where $M$ and $K$ are hyperk\"ahler 4-manifolds in the limit in
which $K$ shrinks to zero size and the space of ``triholomorphic
curves'' (or hyperinstantons) $M\to \eM_{\HH}(K)$.  It seems plausible
that the results in this paper can be used to understand the geometry
of the space of hyperinstantons better.

In analogy to what happens in four dimensions, certain octonionic
instantons can be understood as monopoles in seven dimensions.  These
equations, which generalise the Bogomol'nyi equation, can be defined
on any riemannian 7-manifold $M$ of $G_2$ holonomy.   Very little
is known about the moduli spaces of these monopoles, but it follows
from the results in this paper that when $M$ is flat, the moduli space
is OK.

\section*{Acknowledgement}
It is a pleasure to thank Bobby Acharya, Krzysztof Galicki, George
Papadopoulos, Sonia Stanciu and Jim Stasheff for conversations, Chris
Hull for also telling me about his results \cite{ChrisYM} prior to
publication, and especially Bill Spence for many conversations and
comments on a previous version of this paper.

\appendix

\section{Some octonionic geometry}

In this appendix we summarise the basic notions of octonionic geometry
{\em as used in this paper\/}.  Octonionic geometries and their
torsioned generalisations have been studied recently in \cite{GPS}
where they are exhibited as the geometries of the moduli space of
some solitonic black holes.  As these authors never define the term
octonionic K\"ahler, our definition below is not in conflict with that
paper.  Nevertheless, our definition is rather narrow and in order to
obtain interesting geometries one must relax it, either as was done in
\cite{GPS} or alternatively as we suggest below.

We start with $\RR^8$, which we think of as the octonions $\OO$.  The
real matrices $I^i$ defined by left (or right) multiplication by the
imaginary unit octonions satisfy the Clifford algebra $\Cl(7)$ in
\eqref{eq:cl7}.  In particular, each $I^i$ is complex structure
relative to which the standard euclidean metric is hermitian.

Let $X$ be a riemannian manifold.  For the purposes of this paper, we
will say that $X$ is {\em octonionic (almost) hermitian\/} if it
admits orthogonal (almost) complex structures $\{I^i\}$ satisfying the
algebra \eqref{eq:cl7}. In addition, we will say that an octonionic
hermitian manifold $X$ is {\em octonionic K\"ahler\/} (or {\em OK\/})
if the associated 2-forms $\omega^i$ are K\"ahler.

The existence of such structures on a riemannian manifold imposes
strong constraints on the manifold.  First of all we have that the
dimension of a finite-dimensional octonionic (almost) hermitian
manifold $X$ is divisible by $8$.  This follows from the fact that
each tangent space $T_pX$ admits an action of $\Cl(7)$, whose
representations are always $8k$-dimensional.  The geometry is also
very constrained.  For example, an 8-di\-men\-sion\-al OK manifold $X$
is necessarily flat.  This can be proven as follows.  The fact that
$X$ is OK means that $d\omega^i = \nabla g = 0$ whence $\nabla I^i=0$.
Because the $I^i$ generate $\Cl(7)$, it follows that the holonomy
group commutes with the action of $\Cl(7)$.  Since $\Cl(7)$ acts
irreducibly on the tangent space, the restricted holonomy group is
trivial.  It seems rather likely that the geometry of OK manifolds is
similarly constrained in higher dimensions.

This prompts us to try to generalise OK geometry in such a way that it
admits interesting examples.  For example, as done in \cite{GPS} one
can relax the condition that $\nabla$ be torsionless and also
substitute $\nabla I^i = 0$ for a weaker condition (see (3.33) in
\cite{GPS}).  This yields the so-called OKT geometries.  Another
approach, more in line with quaternionic K\"ahler geometry, would be
to demand that the almost complex structures $I^i$, satisfying
\eqref{eq:cl7}, only exist locally.  Then one would impose that the
7-dimensional subbundle of the 2-forms spanned by the $I^i$, instead
of being trivial, be preserved by the holonomy group.  In eight
dimensions, as we saw in Section~\ref{sec:mfds}, this singles out
those riemannian manifolds whose holonomy group is contained in
$\Spin(7)$.   In $8k \geq 16$ dimensions any such manifold must be
reducible, as can be gleaned from Berger's list of irreducible
holonomy representations.  Nevertheless it seems interesting to try
and develop a theory of such manifolds and in particular to try to use
them to construct 8-dimensional $\Spin(7)$ holonomy manifolds by a
quotient construction.  Additionally one could also relax the
torsionless condition on $\nabla$, and investigate quotient
constructions there.  We hope to turn to some of these topics in the
near future.

%

\begin{thebibliography}{10}

\bibitem{AFOS}
BS~Acharya, JM~Figueroa-O'Farrill, M~O'Loughlin, and B~Spence, \emph{Euclidean
  {D}-branes and higher dimensional gauge theory}, {\tt hep-th/9707118}.

\bibitem{AOS}
BS~Acharya, M~O'Loughlin, and B~Spence, \emph{Higher dimensional analogues of
  {D}onaldson--{W}itten theory}, Nucl. Phys. \textbf{B503} (1997), 657, {\tt
  hep-th/9705138}.

\bibitem{Matrix}
T~Banks, W~Fischler, SH~Shenker, and L~Susskind, \emph{M theory as a matrix
  model: a conjecture}, Phys. Rev. \textbf{D55} (1997), 5112, {\tt
  hep-th/9610043}.

\bibitem{BKS2}
L~Baulieu, H~Kanno, and IM~Singer, \emph{Cohomological {Y}ang--{M}ills theory
  in eight dimensions}, {\tt hep-th/9705127}.

\bibitem{BKS}
L~Baulieu, H~Kanno, and IM~Singer, \emph{Special quantum field theories in eight and other dimensions},
  {\tt hep-th/9704167}.

\bibitem{MatrixT4T5}
M~Berkooz, M~Rozali, and N~Seiberg, \emph{Matrix description of {M}-theory on
  {$T^4$} and {$T^5$}}, {\tt hep-th/9704089}.

\bibitem{BTESYM}
M~Blau and G~Thompson, \emph{Euclidean {SYM} theories by time reduction and
  special holonomy manifolds}, {\tt hep-th/9706225}.

\bibitem{CDFN}
E~Corrigan, C~Devchand, DB~Fairlie, and J~Nuyts, \emph{First-order equations
  for gauge fields in spaces of dimension greater than four}, Nucl. Phys.
  \textbf{B214} (1983), 452--464.

\bibitem{DT}
SK~Donaldson and RP~Thomas, \emph{Gauge theory in higher dimensions}, Oxford
  Preprint, 1996.

\bibitem{FaNu}
DB~Fairlie and J~Nuyts, \emph{Spherically symmetric solutions of gauge theories
  in eight dimensions}, J. Phys. A: Math. Gen. \textbf{17} (1984), 2867--2872.

\bibitem{FIM}
JM~Figueroa-O'Farrill, A~Imaanpur, and J~McCarthy, \emph{Supersymmetry and
  gauge theory in {C}alabi--{Y}au 3-folds}, {\tt hep-th/9709178}.

\bibitem{FKS2}
JM~Figueroa-O'Farrill, C~K{\"o}hl, and B~Spence, \emph{Supersymmetric
  {Y}ang--{M}ills, octonionic instantons and triholomorphic curves}, {\tt
  hep-th/9710082}.

\bibitem{FuNi}
S~Fubini and H~Nicolai, \emph{The octonionic instanton}, Phys. Lett.
  \textbf{155B} (1985), 369--372.

\bibitem{GPS}
G~Gibbons, G~Papadopoulos, and K~Stelle, \emph{{HKT} and {OKT} geometries on
  soliton black hole moduli spaces}, {\tt hep-th/9706207}.

\bibitem{ChrisYM}
CM~Hull, \emph{Higher dimensional {Y}ang--{M}ills theories and topological
  terms}, {\tt hep-th/9710165}.

\bibitem{IP}
TA~Ivanova and AD~Popov, \emph{Some solutions of the {Y}ang--{M}ills equations
  in dimension greater than four}, J. Math. Phys. \textbf{34} (1993), 674--680.

\bibitem{ReyesCarrion}
R~{Reyes Carri\'on}, \emph{A generalization of the notion of instanton}, to
  appear in Differential Geometry and its applications.

\bibitem{RCPhD}
R~{Reyes Carri\'on}, \emph{Some special geometries defined by {L}ie groups}, Ph.D. thesis,
  Oxford University, 1993.

\bibitem{SeibergT5}
N~Seiberg, \emph{New theories in six dimensions and matrix description of
  {M}-theory on {$T^5$} and {$T^5/\mathbb{Z}_2$}}, {\tt hep-th/9705221}.

\bibitem{Thomas}
RP~Thomas, \emph{Gauge theory on {C}alabi--{Y}au manifolds}, Ph.D. thesis,
  Oxford University, 1997.

\bibitem{Ward}
RS~Ward, \emph{Completely solvable gauge-field equations in dimension greater
  than four}, Nucl. Phys. \textbf{B236} (1984), 381--396.

\end{thebibliography}

%
\end{document}